\documentclass[aps,pra,twocolumn,superscriptaddress,groupedaddress]{revtex4}
\usepackage{amssymb}
\usepackage{cancel}
\usepackage{color,graphicx}
\usepackage{amsmath}
\usepackage{amsbsy}
\usepackage{amsthm}
\usepackage{bbm}
\usepackage{epsfig}
\usepackage{lscape}
\usepackage{float}
\usepackage{graphicx}
\usepackage{subfigure}
\usepackage{dcolumn}
\usepackage{bbm}
\usepackage{color,epstopdf}
\usepackage{amscd}
\usepackage{amsfonts}  
\usepackage[]{amsmath}
\usepackage{amssymb}    
\usepackage{mathrsfs}
\usepackage{verbatim}
\usepackage[]{cases}
\usepackage{amsmath}
\usepackage{wasysym}
\usepackage[utf8]{inputenc}
\usepackage[T1]{fontenc}
\usepackage{mathtools}

\newcommand{\mcdots}{\cdot\!\cdot\!\cdot}

\newcommand{\ket}[1]{\left| {#1} \right\rangle}
\newcommand{\bra}[1]{\left\langle {#1}\right|}
\newcommand{\braket}[1]{\left\langle {#1} \right\rangle}
\newcommand{\abs}[1]{\left| \,{#1} \right|}

\newcommand{\se}{\text{\tiny{$\mathcal{SE}$}}}
\newcommand{\s}{\text{\tiny{$\mathcal{S}$}}}
\newcommand{\e}{\text{\tiny{$\mathcal{E}$}}}

\newcommand\ro{\hat\rho}

\newcommand\Vo{\hat V}
\newcommand\Ao{\hat A}

\newcommand\phio{\hat\phi}

\newcommand\fo{\hat f}

\newcommand\Tr{\mathrm{Tr}}

\makeatletter
\newcommand{\oset}[2]{
   {\mathop{#2}\limits^{\vbox to -.5\ex@{\kern-\tw@\ex@
    \hbox{\scriptsize #1}\vss}}}}
\makeatother

\newcommand\harpr[1]{\mathstrut\mkern2.5mu#1\mkern-11mu\raise1.5ex%
  \hbox{$\scriptscriptstyle\rightharpoonup$}}

\newcommand\harpl[1]{\mathstrut\mkern2.5mu#1\mkern-11mu\raise1.5ex%
  \hbox{$\scriptscriptstyle\leftharpoonup$}}

\begin{document}
\title{Stochastic unravellings of non-Markovian completely positive \\and trace preserving maps}
\author{G. Gasbarri}
\email{giulio.gasbarri@ts.infn.it}
\affiliation{Department of Physics, University of Trieste, Strada costiera 11, 34151 Trieste, Italy}  
\author{L. Ferialdi}
\email{l.ferialdi@soton.ac.uk}
\affiliation{Department of Physics, University of Ljubljana, Jadranska 19, SI-1000 Ljubljana, Slovenia}
\affiliation{Department of Physics and Astronomy, University of Southampton, SO17 1BJ, United Kingdom}
\date{\today}
\begin{abstract}
We consider open quantum systems with factorized initial states, providing the structure of the reduced system dynamics, in terms of environment cumulants. We show that such completely positive (CP) and trace preserving (TP) maps can be unraveled by linear stochastic Schr\"odinger equations (SSEs) characterized by 
 sets of colored stochastic processes (with $n$-th order cumulants). We obtain both the conditions such that the SSEs provide CPTP dynamics, and those for unraveling an open system dynamics.
We then focus on Gaussian non-Markovian unravellings, whose known structure displays a functional derivative. We provide a novel description that replaces the functional derivative with a recursive operatorial structure. Moreover, for the family of quadratic bosonic Hamiltonians, we are able to provide an explicit operatorial dependence for the unravelling.
\end{abstract}

\maketitle

\section{Introduction}
 Stochastic Schr\"odinger equations (SSEs) were introduced in the framework of measurement theory,  where the effect of repeated measurements is described by the introduction of a Wiener process in the Schr\"odinger equation~\cite{cmt}. In the following years, different SSEs were proposed in the context of collapse models where one modifies the Schr\"odinger dynamics in such a way to account for the wave packet reduction within a unique dynamical principle~\cite{cm}.  
Markovian SSEs, i.e. those displaying white noises, have been deeply investigated thanks to the possibility of exploiting Ito stochastic calculus \cite{monitoringM,VanKamp,Barchielli}.
The solutions of these SSEs ``unravel'' (i.e. reproduce in average) Markovian open systems dynamics. However, many physical systems cannot be described by Markovian dynamics, and they need non-Markovian ones~\cite{NMsys,deVAlo17}. This fact pushed the development and investigation of non-Markovian SSEs.
The extension to non-Markovian quantum state diffusion was first tackled by Diosi and Strunz in~\cite{DioStr97}, where Gaussian, complex, colored stochastic processes were considered. These seminal works were followed by others both in the realm of collapse models~\cite{AdlBas,BasFer09,FerBas12,disscm}, and in that of continuous measurement~\cite{monitoringNM} (for a more detailed historical review see~\cite{deVAlo17}).  We should mention that in this context one considers stochastic equations with discontinuous or jump processes, both for Markovian and non-Markovian dynamics~\cite{cm,barchetta}. We are instead interested in stochastic processes that are continuous and continuously differentiable, and we will focus on these in the remainder of the paper. SSEs are also exploited, as method alternative to master equations, to numerically investigate open quantum systems. The advantage is that SSEs typically require less numerical resources than master equations (because the wave functions Hilbert spaces have smaller dimensionality than those of density matrices). Different expansions methods have been proposed to investigate non-Markovian SSEs, but only in few cases these equations  have been analytically solved~\cite{BasFer09,FerBas12,nll,ansol}. 
We should also mention the original method proposed by Tilloy~\cite{Til17}, who introduces new stochastic terms in order to rewrite non-Markovian SSEs as averages over explicit time-local (bi-)stochastic differential equations.
Recently, the first full characterization of Gaussian linear stochastic unravellings have been provided in~\cite{DioFer14}. Such a characterization is implicit, as it depends on a functional derivative (further explained in Sec.IV).
 The symmetries of these unravellings were analyzed in~\cite{Bud15}, while a measurement interpretation in terms of Bargmann states has been provided in~\cite{Megetal17}.
 
Most of the results on non-Markovian SSEs in the literature concern Gaussian stochastic processes, 
and only few papers attempt to use continuous, non-Gaussian stochastic processes in very specific cases~\cite{nngaus}. 
The aim of this paper is to take a step forward in the theoretical description of non-Markovian unravelings. On the one side, we show that it is possible to unravel non-Gaussian, completely positive (CP) and trace preserving (TP) non-Markovian open system dynamics, and we provide the condition for the existence of such an unraveling (Sec.III). In general, if the open system dynamics is non-Gaussian, the SSE will display continuous colored stochastic processes characterized by their n-th order cumulants. On the other hand, we provide a novel description of Gaussian non-Markovian unravelings that does not make use of the functional derivative and explicitly depends on the system operators (Sec.IV). In the next section we start by we investigating the structure of (non-Gaussian) CPTP open system dynamics, that is the reference dynamics that we want to unravel.

\section{Open Quantum Systems Map}
We consider a system ($\mathcal{S}$) interacting with a generic environment ($\mathcal{E}$). 
The evolution of the open system density matrix $\ro_{\se}$ in the interaction picture is described by the Von-Neumann equation ($\hbar=1$)
\begin{align}\label{vn}
i \partial_{t} \hat{\rho}_{\se}(t) = \left[\hat{V}_{t},\hat{\rho}_{\se}(t)\right]\,,
\end{align}
where $\Vo_t$ is a generic interaction Hamiltonian between the system and the environment, that can be rewritten in total generality as: 
\begin{align}\label{v}
\hat{V}_{t}=\sum_\alpha\hat{f}_{\alpha}(t)\hat{\phi}_{\alpha}(t)\,,
\end{align}
where $\hat{f}^{\alpha}(t)$ and $\phio_{\alpha}(t)$ respectively are Hermitian system and environment operators.
The solution of Eq.~\eqref{vn} can be formally written as:
\begin{align}\label{eq:3}
\rho_{\se}(t)=\hat{\Phi}_{t}\rho_{\se}\hat{\Phi}^{\dagger}_{t}
\end{align}
with
\begin{align}\label{eq:phi0}
\hat{\Phi}_{t}= \mathcal{T}\exp\left[-i\int_{0}^{t}d\tau\, \sum_\alpha\hat{f}_{\alpha}(\tau){\hat{\phi}}_{\alpha}(\tau)\right]\,.
\end{align}
The time ordering $\mathcal{T}$ acts on a generic function of the type $f(\int_{0}^{t}d\tau\hat{ V}_{\tau})$, by ordering the products of operators $\hat{V}_{\tau}$ of its Taylor expansion:
\begin{align}\label{eq:taylort}
\mathcal{T}f\left(\int_{0}^{t}d\tau\hat{V}_{\tau}\right)= \sum_{n=0}^{\infty}\frac{1}{n!}\mathcal{T}\left[\left(\int_{0}^{t}d\tau\hat{V}_{\tau}\right)^{n}\right]\frac{\partial^{n}f(x)}{\partial{x}^n}\bigg|_{x=0}\,,
\end{align}
provided the series exists.
Since we are interested in the effective evolution of the system $\mathcal{S}$ we trace over the environment degrees of freedom to obtain:
\begin{align}\label{eq:m0tilde}
\hat{\rho}_{\s}(t)= \Tr_{\e}[\hat{\Phi}_{t}\,\hat{\rho}_{\se} \, \hat{\Phi}_{t}^{\dagger}]\,.
\end{align}
Under the assumption of factorized initial state $\hat{\rho}_{\se}=\hat{\rho}_{\s}\otimes\hat{\rho}_{\e}$, the effective evolution can be described by the action of a dynamical map $\widetilde{\mathcal{M}}_{t}$ on the system initial state $\rho_{s}$.
By replacing Eq.~\eqref{eq:phi0} in Eq.~\eqref{eq:m0tilde}, one can write the dynamical map as: 
\begin{align}\label{eq50}
\widetilde{\mathcal{M}}_{t}[\hat{\rho}_{\s}]\!=\!\braket{\mathcal{T}e^{-\frac{i}{2}\int_{0}^{t}\!\!d\tau \sum_\alpha [{f}_{\alpha}^{+}(\tau)\phi^{-}_{\alpha}(\tau)+{f}_{\alpha}^{-}(\tau){\phi}^{+}_{\alpha}(\tau)]}\hat{\rho}_{\s}}
\end{align}
where $\braket{\dots}=\Tr_{\e}[\dots \hat{\rho}_{\e}]$, and the superscripts $+/-$ denote the super-operators $f^{\pm} \hat{\rho}= \hat{f}\hat{\rho}\pm\hat{\rho}\hat{f}$ (similar definitions hold for the super-operators $\phi^{\pm}$).
We rewrite the map $\widetilde{\mathcal{M}}_{t}$ as a time ordered exponential with a time dependent generator, by means of a cumulant expansion (see Appendix A for explicit calculation):
\begin{align}\label{cuMt}
\widetilde{\mathcal{M}}_{t}[\hat{\rho}_{\s}]=\mathcal{T}\exp\left[\sum_{n=1}^{\infty}(-i)^{n}\int_{0}^{t}d\boldsymbol{\tau}_n\,\tilde{k}_{n}(\boldsymbol{\tau}_n)\right]\hat{\rho}_{\s}\,,
\end{align}
where $\boldsymbol{\tau}_n=(\tau_{1},\cdots \tau_{n})$ ($n$ denoting the length of the array). The terms of the expansion are defined as
\begin{align}\label{kn0}
\tilde{k}_{n}(\boldsymbol{\tau}_n)&=\sum_{q=0}^{n}\sum_{\alpha_i,\mathcal{P}_{q}}f^{-}_{\alpha_{1}}(\tau_1)f^{l_{2}}_{\alpha_{2}}(\tau_2)\mcdots f^{l_{n}}_{\alpha_{n}}(\tau_n)\,\widetilde{C}^{+\bar{l}_2\cdots \bar{l}_{n}}_{\alpha_{1}\cdots \alpha_{n}}\!(\boldsymbol{\tau}_n)\,,
\end{align}
where $\widetilde{C}$ are the ordered bath cumulants defined in Eq.~\eqref{cumC}. We used the following notation: $l_{i} \in \{+, -\}$, $\bar{l}_i=-l_i$, $\mathcal{P}_{q}$ is the permutation of the indexes $l_{i}$ such that $q$ is the number of plus signs in the array $({l}_2,\cdots, {l}_n)$, and $\alpha_{i}$ denotes the sum over all $\{\alpha_{1},\cdots,\alpha_{n}\}$.
We stress that the first super-operator on the left in $\tilde{k}_{n}(\boldsymbol{\tau}_{n})$ is always $f^{-}$. This is an important feature because it guarantees that the map is trace preserving (indeed $\Tr_{\s}[f^{-}\hat{O}\hat{\rho}_{s}]=0$). Note moreover that the time ordering acts on the basic elements $f$, hence one has first to replace Eq.~\eqref{kn0} in Eq.~\eqref{cuMt} , and then apply $\mathcal{T}$.

We eventually notice that Eq.~\eqref{cuMt} recovers the one derived by Diosi and  Ferialdi~\cite{DioFer14}  when the (Gaussian) environment is completely characterized by its second cumulant.
Indeed, in such a case Eq.~\eqref{cuMt} reduces to
\begin{align}
&\widetilde{\mathcal{M}}_{t}=\mathcal{T}\exp\Bigg[-\int_{0}^{t}d\boldsymbol{\tau}_{2}\sum_{\alpha_{1},\alpha_{2}}f^{-}_{\alpha_{1}}(\tau_{1})\times\\
&\hspace{2.2cm}\times\Big(f^{-}_{\alpha_{2}}(\tau_{2})\widetilde{C}^{++}_{\alpha_{1}\alpha_{2}}(\boldsymbol{\tau}_{2})+f^{+}\widetilde{C}^{+-}_{\alpha_{1}\alpha_{2}}(\boldsymbol{\tau}_{2})\Big)\Bigg]\,,\nonumber
\end{align}
that coincides with Eq.(17) in~\cite{DioFer14}.

\section{General unravelling} 

We now  consider a stochastic Schr\"{o}dinger equation in interaction picture 
\begin{align}\label{eq:stoch}
i\partial_{t}\ket{\psi_{t}}=\hat{V}_{t}\ket{\psi_{t}}.
\end{align}
where  $\hat{V}_{t}$ is an operator valued stochastic field in the Hilbert space $\mathcal{H}
$, that can be decomposed as a sum of Hermitian system operators $\hat{f}_{\alpha}$ like in Eq.~\eqref{v}, where the operators $\hat{\phi}_{\alpha}$ are replaced by a set of complex stochastic processes $\{\phi_{\alpha}\}$.
A decomposition with non-Hermitian operators can be obtained performing a unitary transformation as shown in \cite{Bud15}.
Unlike previous works on non-Markovian stochastic unravellings, we consider general stochastic processes, that are completely characterized by their $n$-th order correlation functions, with $n$ arbitrary.
Equation~\eqref{eq:stoch} is solved by Eq.~\eqref{eq:phi0} with $\hat{\phi}\rightarrow\phi$, and the average dynamics is described by the map $\mathcal{M}_{t}$, defined as follows:
\begin{align}\label{eq:m}
\mathcal{M}_{t}[\hat{\rho}_\s]= \braket{\hat{\Phi}_{t}\,\hat{\rho}_\s \, \hat{\Phi}_{t}^{\dagger}}\,
\end{align}
where $\hat{\rho}_\s=\ket{\psi_{0}}\bra{\psi_{0}}$ is a pure initial state, and now $\braket{\cdots}$ denotes the stochastic average.
In order for $\mathcal{M}_t$ to be a physical map, it must be CPTP. The definition~\eqref{eq:m} itself can be understood as the Kraus-Stinespring decomposition of $\mathcal{M}_{t}$, guaranteeing its CP~\cite{Sti55}. By replacing Eq.~\eqref{eq:phi0} in Eq.~\eqref{eq:m}, one can conveniently rewrite $\mathcal{M}_t$ like in Eq.~\eqref{eq50},
where now $\phi^{\pm}=\phi\pm\phi^{*}$ are twice the real and imaginary parts of $\phi$.
 In order to find the conditions under which the averaged map is TP, we perform --analogously to the quantum case--  a cumulant expansion (see Appendix \ref{app:cum}), obtaining Eq.~\eqref{cuMt} where $\tilde{k}$ is replaced by
\begin{align}\label{kn}
k_{n}(\boldsymbol{\tau}_n)&=\sum_{q=0}^{n}\sum_{\alpha_i,\mathcal{P}_{q}}f^{l_{1}}_{\alpha_{1}}(\tau_1)\mcdots f^{l_{n}}_{\alpha_{n}}(\tau_n)\,C^{\bar{l}_{1}\cdots \bar{l}_{n}}_{\alpha_{1}\cdots \alpha_{n}}\!(\boldsymbol{\tau}_n)\,.
\end{align}
$C^{\bar{l}_{1}\cdots \bar{l}_{n}}_{\alpha_{1}\cdots \alpha_{n}}\!(\boldsymbol{\tau}_n)$ are the $n$-th order cumulants of the set of stochastic processes $\phi_\alpha$ (see Appendix \ref{app:cum} for the explicit expression). It is important to stress that the stochastic cumulants $C$ in principle differ from the quantum cumulants $\widetilde{C}$. This is due to the fact that the trace over quantum degrees of freedom in general cannot be described as a stochastic average.

We perform the trace of $\mathcal{M}_t$, and we observe that the contributions of the type $f^{-}_{\alpha_{1}}(\tau_1)\cdots f^{l_{n}}_{\alpha_{n}}(\tau_n)C^{+\cdots \bar{l}_{n}}_{\alpha_{1}\cdots \alpha_{n}}\!(\boldsymbol{\tau}_n)$ (i.e. those with $l_1=-$) in each $k_n(\boldsymbol{\tau}_{n})$ are killed by the cyclicity of the trace (remember that $f^-$ is a commutator, and accordingly $\Tr\{f^-\hat{O}\}=0$):
\begin{align}\label{eq:tr}
\!\Tr\!\left\{\mathcal{M}_{t}[\hat{\rho}_\s]\right\}\!=\!& \Tr\left\{\mathcal{T}\exp\left[\int_0^td\boldsymbol{\tau}_n\sum_{n=1}^{\infty}\sum_{q=1}^{n}\sum_{\alpha_i,\mathcal{P}_{q}}(-i)^n\nonumber\right.\right.\\
&f^{+}_{\alpha_1}(\tau_{1})f^{l_{2}}_{\alpha_2}(\tau_{2})\mcdots f^{l_{n}}_{\alpha_n}(\tau_{n})\,C^{- \bar{l}_{2}\cdots \bar{l}_{n}}_{\alpha_1\cdots\alpha_n}(\boldsymbol{\tau}_n)\Bigg]\hat{\rho}_\s\Bigg\}.
\end{align}
Recalling that $\mathcal{M}_t$ is TP if $\Tr\left\{\mathcal{M}_{t}[\hat{\rho}_\s]\right\}=\Tr\left\{\hat{\rho}_\s\right\}$, one finds that this condition is satisfied only if the exponent in the previous equation is zero, i.e. if either the series or each of its terms are zero. In the first case one needs to fine-tune the stochastic cumulants $C$ (by manipulating the processes $\phi_\alpha$) in such a way that the series in the exponent sum to zero. This is in general a very difficult task and we are not aware of any technique that allows to provide an explicit  constraint.
The second option, though being a more stringent requirement, allows to find and explicit constraint:
\begin{align}\label{condizioneC}
C^{- \bar{l}_{2}\cdots \bar{l}_{n}}_{\alpha_1\cdots\alpha_n}(\boldsymbol{\tau}_n)=0\,,\qquad\forall\, n\,,
\end{align}
 i.e. all the cumulants obtained tracing at least one purely imaginary noise $\phi^{-}$ are zero. Accordingly, the average map $\mathcal{M}_t$ is TP if it is generated by an interaction potential $\hat{V}(t)$ that displays only real stochastic processes $\phi_\alpha$. In this case $\hat{V}(t)$ is purely Hermitian, and  Eq.~\eqref{kn} is replaced by ($\phi_\alpha=\phi_\alpha^+$)
\begin{align}
k_{n}(\boldsymbol{\tau}_n)= \sum_{\alpha_{i}}f^{-}_{\alpha_1}(\tau_{1})\mcdots f^{-}_{\alpha_n}(\tau_{n})\,C^{+\cdots +}_{\alpha_1\cdots\alpha_n}(\boldsymbol{\tau}_n)\,.
\end{align}
However, it is well known that a SSE of this kind generates a dynamics that cannot describe dissipative phenomena~\cite{BasFer09}. Dissipative dynamics are indeed unraveled by SSEs  displaying an interaction potential~\eqref{v} with complex stochastic processes~\cite{FerBas12,BasIppVac05} (provided that the system operators are Hermitian). Since the previous calculations explicitly show that the unravelling~\eqref{eq:stoch} with complex noises leads to a dynamics that is not TP, we now investigate whether it is possible to obtain such a TP map starting from a SSE different from~\eqref{eq:stoch}. We do so by adding new terms to the stochastic map $\hat{\Phi}_t$, and we seek the conditions on these new terms such that the average map is TP.
We stress that one might make $\mathcal{M}_t$ trace preserving by modifying directly the $k_n(\boldsymbol{\tau}_{n})$ of Eq.~\eqref{kn}. This operation, though leading to a TP map, does not guarantee that $\mathcal{M}_t$ preserves the factorized structure of Eq.~\eqref{eq:m}, which is an essential requirement to obtain a SSE unravelling it.


The most general modification of $\hat{\Phi}_{t}$ is obtained by adding to the exponent an operator functional $g(\int_0^t d\tau\hat{f}_\tau)$, that for later convenience we Taylor expand around $\int_{0}^{t}d\tau \hat{f}_{\tau}=0$ (see e.g. Appendix A of \cite{abc}).
This leads to a new map $\hat{\Xi}_t$:
\begin{equation}\label{xi}
\hat{\Xi}_{t}\!=\!\mathcal{T}\!\exp\!\left[\!-i\!\!\int_{0}^{t}\!\!d\tau \! \sum_\alpha\hat{f}_{\alpha}(\tau){\phi}_{\alpha}(\tau)\!-\!\sum_{n=1}^{\infty}\int_{0}^{t} \!\!d\boldsymbol{\tau}_{n}\hat{O}_n\!(\boldsymbol{\tau}_{n})\!\right],
\end{equation}
where the operators $\hat{O}_n(\boldsymbol{\tau}_{n})$ are defined by
\begin{align}
\int_{0}^{t} \!\!d\boldsymbol{\tau}_{n}\hat{O}_n\!(\boldsymbol{\tau}_{n}) = \frac{1}{n!}\left(\int_{0}^{t}d\tau\hat{f}_{\tau}\right)^{n}\frac{\partial^{n}g(x)}{\partial{x}^n}\bigg|_{x=0}.
\end{align} 
We  stress that $\hat{O}_n(\boldsymbol{\tau}_{n})$ are deterministic: if they were stochastic they would lead to a new map that could be rewritten in the form~\eqref{eq:phi0} (with new stochastic processes), therefore not solving the trace issue.
The stochastic map~\eqref{xi} generates the average dynamics $\mathcal{N}_{t}[\hat{\rho}_\s]= \langle\hat{\Xi}_{t}\,\hat{\rho}_\s \, \hat{\Xi}_{t}^{\dagger}\rangle$. We remark that this dynamics is CP even if $\hat{\Xi}_{t}$ is truncated at the $n$-th order, because the Kraus-Stinespring structure of Eq.~\eqref{eq:m} is preserved.
Exploiting the identity $\mathcal{T}e^{i(a+b)}=\mathcal{T}e^{i\mathcal{T}log(e^{ia})+b}$ and the calculations in Appendix \ref{app:cum}, one finds that
\begin{align}\label{mtilde}
\mathcal{N}_{t} &= \mathcal{T}\exp\left[\sum_{n=1}^{\infty} \int_0^t d\boldsymbol{\tau}_{n}\Big((-i)^{n}k_{n}(\boldsymbol{\tau}_{n})-\mathcal{O}_{n}(\boldsymbol{\tau}_{n})\Big)\right]
\end{align}
with 
\begin{align}\label{on}
\mathcal{O}_{n}(\boldsymbol{\tau}_n)&= H_{n}^{+}(\boldsymbol{\tau}_{n})+A^{-}_{n}(\boldsymbol{\tau}_{n})\,.\end{align}
The super-operators $H_{n}^{+}$, $A_{n}^{-}$ are built --with the known rules-- respectively from the Hemitian and anti-Hermitian parts of $\hat{O}_n$: $\hat{H}_{n}=(\hat{O}_n+ \hat{O}_n^{\dagger})/2$ and $\hat{A}_n=(\hat{O}_n- \hat{O}_n^{\dagger})/2$.
By performing the trace of Eq.~\eqref{mtilde}, one finds that the averaged map $\mathcal{N}_{t}$ is trace preserving if the following condition is satisfied (see Appendix~\ref{app:tr}):
\begin{align}\label{eq:Aop}
 &\hat{A}_{n}(\boldsymbol{\tau}_{n})=\nonumber\\
&(-i)^{n}\sum_{q=1}^{n}\sum_{\alpha_{i},\mathcal{P}_{q}}\hat{f}_{\alpha_{1}}(\tau_{1})\harpl{f}^{l_{2}}_{\alpha_{2}}(\tau_{2})\!\cdots \harpl{f}^{l_{n}}_{\alpha_{n}}(\tau_{n})C^{-\bar{l}_{2}\cdots \bar{l}_{n}}_{\alpha_{1}\!\cdots \alpha_{n}}\!(\boldsymbol{\tau}_n),
\end{align}
where the super-operators $\harpl{f}^{\pm}$ are defined as $\hat{O}\harpl{f}^{\pm}=\hat{O}\hat{f}\pm\hat{f}\hat{O}$, the arrow denoting the fact that these super-operators are acting on the left.
We recall that $\bar{l}_{1}= -$ because of the trace operation (see Eq.~\eqref{eq:tr}). 
We then find that the request of $\mathcal{N}_{t}$ being TP map can be satisfied --for a generic noise-- if a multi-time anti-Hermitian operator of the form in Eq.~\eqref{eq:Aop} is added to the map $\hat{\Xi}_{t}$.
Accordingly, by replacing Eq.~\eqref{eq:Aop} in Eq.~\eqref{xi} we obtain the structure of a general stochastic map that generates an average CPTP dynamics.
Interestingly, the fact that only the anti-Hermitian $\hat{A}_n$ are involved in the TP condition implies that the Hermitian contributions $\hat{H}_{n}$ can be chosen freely. This degree of freedom can be exploited to tune the average dynamics $\mathcal{N}_t$ obtained from a stochastic evolution, in such a way that it recovers an open system  dynamics $\widetilde{\mathcal{M}}_t$. 
 By matching the exponents of Eq.~\eqref{mtilde} and Eq.~\eqref{cuMt}, one finds that $\hat{\Xi}_{t}$ unravels $\widetilde{\mathcal{M}}_t$ if
\begin{align}
H_{n}^{+}(\boldsymbol{\tau}_{n})&= (-i)^n \left(k_{n}(\boldsymbol{\tau}_{n})-\tilde{k}_{n}(\boldsymbol{\tau}_{n})\right)-A^{-}_{n}(\boldsymbol{\tau}_{n})\,.
\end{align}
One might question whether it is possible to find a more explicit expression for this formal equation, e.g. by decomposing the operator $\hat{O}_n$ over the set $\{{\hat{f}_\alpha}\}$ in Eq.~\eqref{xi}. Unfortunately, such a decomposition is helpful only in the Gaussian case. The calculations showing this are rather instructive and we reported them in Appendix~\ref{app:tr}. As a matter of facts, there are two cases for which one can obtain a simpler expression for $\hat{H}_{n}$: when all operators $\hat{f}_{\alpha}(t)$ commute at any time ($[\hat{f}_{\alpha}(t),\hat{f}_{\alpha}(\tau)]=0$), and the Gaussian case (stochastic processes characterized only by their first two cumulants). The first case is trivially solved and we will not report it here, while the Gaussian case is more interesting and it will be considered it in the next section.

\section{Gaussian unravelling} 
We consider complex Gaussian stochastic processes, i.e. those completely characterized by their first two cumulants:
\begin{align}
 &\langle{\phi^\pm_{\alpha}(t)\rangle}=2C_\alpha^\pm(t)\nonumber\\
  &\langle{\phi^\pm_{\alpha}(t)^{*}\phi^\pm_{\beta}(\tau)\rangle}- \braket{\phi^\pm_{\alpha}(t)^*}\langle{\phi^\pm_{\beta}(\tau)\rangle}= 4C^{\pm\pm}_{\alpha\beta}(t,\tau)\,.
\end{align}
 The CPTP stochastic map~\eqref{xi} associated to such processes simplifies to
 \begin{align}\label{xigauss0}
\hat{\Xi}_{t}&=\mathcal{T}\exp\!\bigg\{ \!\!-i\!\!\int_{0}^{t} \!\!d\tau_{1}\Big[\hat{f}^{\alpha}(\tau_{1})\big[{\phi}_{\alpha}(\tau_{1})\!-\!C_{\alpha}^{-}(\tau_{1})\big] \nonumber\\
&+\int_{0}^{t} \!\!\!d\boldsymbol{\tau}_2 \hat{f}^{\alpha}(\tau_1)\!\!\left[\!\harpl{f}^{\beta +}\!(\tau_2)C_{\alpha\beta}^{--}(\boldsymbol{\tau}_2)\!+\!\harpl{f}^{\beta -}\!(\tau_2)C_{\alpha\beta}^{-+}(\boldsymbol{\tau}_2)\right]\nonumber\\
&-\int_{0}^{t}d{\tau}_{1} \hat{H}_1({\tau}_{1})-  \int_{0}^{t} d\boldsymbol{\tau}_2\hat{H}_2(\boldsymbol{\tau}_2)\bigg\},
\end{align}
where we are now making use of the Einstein convention of summing over repeated indexes.
In this specific  case we can set $\hat{H}_1(\tau)=0$, and determine the operator $\hat{H}_2(\boldsymbol{\tau}_2)$ in such a way that the average map reproduces the effective Gaussian map generated by the trace over an environment:
 \begin{align}
&\hat{H}_2(\boldsymbol{\tau}_2)= \hat{f}^{\alpha}(\tau_1)\left[\harpl{f}^{-\beta}(\tau_2)C_{\alpha\beta}^{--}(\boldsymbol{\tau}_2)+\harpl{f}^{+\beta}(\tau_2)C_{\alpha\beta}^{-+}(\boldsymbol{\tau}_2)\right].
 \end{align} 
Replacing this equation in Eq.~\eqref{xigauss} we obtain the following expression for the stochastic map
\begin{align}\label{xigauss}
\hat{\Xi}_{t}=&\mathcal{T}\exp\left\{-i\int_0^t d\tau\hat{f}^{\alpha}(\tau)\left({\phi}_{\alpha}(\tau)-C^{-}_{\alpha}(\tau)\right)\right.\\
&\left.+\int_0^t d\boldsymbol{\tau}_2 \hat{f}^{\alpha}(\tau_1)\hat{f}^{\beta}(\tau_2)\left(C_{\alpha\beta}^{--}(\boldsymbol{\tau}_2)+C_{\alpha\beta}^{-+}(\boldsymbol{\tau}_2)\right)\right\}.\nonumber
\end{align}
Since $C^{-}_{\alpha}(\tau)$ in~\eqref{xigauss} simply gives a shift of the mean value of  $\phi_{\alpha}$, we absorb it as follows $\phi_{\alpha}(\tau)-C^{-}_{\alpha}(\tau)\to\phi_{\alpha}(\tau)$, in order to simplify the notation.
This leads to
\begin{align}\label{xisimp}
\hat{\Xi}_{t}=\mathcal{T}\left\{e^{-i\int_{0}^{t}d\tau \hat{f}^{\alpha}(\tau)\left[{\phi}_{\alpha}(\tau)
- i\int_0^{t} d\tau_1 \hat{f}^{\beta}(\tau_1)K_{\alpha\beta}(\tau,\tau_1)\right]}\right\}\,,
\end{align}
with the following conditions on the new $\phi$:
\begin{align}\label{eq:corK2}
 &\braket{\phi_{\alpha}(t)}=\braket{\phi_{\alpha}(t)^{*}}\nonumber\\
  &K_{\alpha\beta}(t,\tau_1)=\frac{1}{2}\big(\!\braket{\phi_{\alpha}(t)^{*}\phi_{\beta}(\tau_1)}- \braket{\phi_{\alpha}(t)\phi_{\beta}(\tau_1)}\!\big)\theta_{t\tau_1}\,.
 \end{align}
Equation~\eqref{xisimp} is the most general Gaussian, non-Markovian stochastic map that unravels a CPTP dissipative dynamics, and coincides with the one first obtained in~\cite{DioFer14} for $\braket{\phi_{\alpha}}=0$. In order to find the SSE associated to this map, we need to differentiate it with respect to $t$, obtaining
\begin{align}\label{eq:lsd}
\partial_t\hat{\Xi}_t=&-i \hat{f}^{\alpha}(t)\phi_{\alpha}(t)\,\hat{\Xi}_{t}\\
&-\mathcal{T}\!\left[\hat{f}^{\alpha}(t)\!\int_{0}^{t}\!\!d\tau K_{\alpha\beta}(t,\tau)\hat{f}^{\beta}(\tau)\right.\nonumber\\
&\hspace{0.5cm}\left.\times e^{-i\int_{0}^{t}d\tau \hat{f}^{\alpha}(\tau)\left[{\phi}_{\alpha}(\tau)
- i\int_0^{\tau} d\tau_1 \hat{f}^{\beta}(\tau_1)K_{\alpha\beta}(\tau,\tau_1)\right]}
\right]\,\nonumber.
\end{align}
The main issue with this equation is that the operator $\hat{f}^{\beta}(\tau)$ in the second line is entangled with the exponential through the time ordering. The approach most widely used to deal with this issue is to exploit the Furutsu-Novikov theorem~\cite{fn}, that allows to rewrite Eq.~\eqref{eq:lsd} as follows:
\begin{align}\label{fn}
\partial_t\hat{\Xi}_{t}=-i\left[ \hat{f}^{\alpha}(t)\phi_{\alpha}(t)\!-\!i\hat{f}^{\alpha}(t)\!\!\int_{0}^{t}\!\!d\tau K_{\alpha\beta}(t,\tau)\frac{\delta}{\delta\phi_{\beta}(\tau)}\right]\hat{\Xi}_{t}\,,
\end{align}
where  $\delta/\delta{\phi_{\beta}(\tau)}$ denotes a functional derivative.
However, this is just a formal and elegant rewriting of Eq.~\eqref{eq:lsd}, that it is very difficult to exploit for practical purposes~\cite{budini1,ansol,BasFer09,FerBas12}. 
We propose an alternative perturbative scheme, that has the merit of allowing for a recursive definition of the expansion terms for the time local generator.
We rewrite Eq.~\eqref{eq:lsd} as follows
\begin{align}\label{eq22}
\partial_t{\hat{\Xi}}_t=&-i \left[\hat{f}^{\alpha}(t)\phi_{\alpha}(t)+ \hat{\Lambda}_{t}\right]\hat{\Xi}_{t}
\end{align}
where $\hat{\Lambda}_{t}$ is a time-local representation of the non-Markovian  contribution to the dynamics.
When inverted, this equation allows to write a formal expression for $\hat{\Lambda}_{t}$:
\begin{align}\label{Lform}
&\hat{\Lambda}_{t}= \left[i\,\partial_t\hat{\Xi}_{t}- \hat{f}^{\alpha}(t)\phi_{\alpha}(t)\,\hat{\Xi}_{t}\right]\hat{\Xi}_{t}^{-1}\,.
\end{align}
We adopt the same strategy as in~\cite{GasFer17}: we Taylor expand the stochastic map \eqref{xigauss} to obtain
\begin{align}\label{eq:xitayl}
\hat{\Xi}_{t}=&
\sum_{n=0}^{\infty}\sum_{k=0}^{n}\frac{1}{n!k!}\mathcal{T}\bigg\{\left(-i\int_0^t d {\tau}\hat{f}^{\alpha}(\tau)\phi_{\alpha}(\tau)\right)^{n-k}\times\nonumber\\
&\left(-\int_0^t d\boldsymbol{\tau}_{2}\hat{f}^{\alpha}(\tau_{1})\hat{f}^{\beta}(\tau_{2})K_{\alpha\beta}(\boldsymbol{\tau}_{2})\right)^{k}\bigg\},
\end{align}
then after some manipulation and explicitly solving the time ordering (see Appendix \ref{app:B}), we write the map $\hat{\Xi}_{t}$ as follows:
\begin{align}\label{xiexp}
\hat{\Xi}_{t}=\sum_{n=0}^{\infty}(-i)^{n}\hat{\xi}_{n}
\end{align}
 with
\begin{align}\label{qn}
\hat{\xi}_{n}=&\,\int_0^t d\boldsymbol{\tau}_{n} \sum_{\alpha_i}\hat{f}_{\alpha_{1}}(\tau_1)\mcdots \hat{f}_{\alpha_{n}}(\tau_n)\nonumber\\
&\times\sum_{k=1}^{\lfloor n/2\rfloor}\!\!\sum_{\mathcal{P}(a)}K_{\alpha_{1} \alpha_{2}}\!(\tau_{a_1},\tau_{a_2})\mcdots K_{\alpha_{2k-1} \alpha_{2k}}\!(\tau_{a_{2k-1}},\tau_{a_{2k}})\nonumber\\
&\times\phi_{\alpha_{2k+1}}(\tau_{a_{2k+1}})\mcdots\phi_{\alpha_{n}}(\tau_{a_{n}})\theta_{\tau_{a_{2k+1}}\cdots\tau_{a_{n}}}\,,
\end{align}
where $\lfloor n/2 \rfloor$ denotes the floor function of $n/2$ (i.e. the greatest integer less than or equal to $n/2$), and $\mathcal{P}(a)$ is the permutation of the indexes $a_{i} \in \{1,\dots,n\}$.
We also expand the right hand side of Eq.~\eqref{Lform} (see Appendix~\ref{app:B} for detailed calculation.):
\begin{align}\label{lambdaxi}
\left[i\,\partial_t\hat{\Xi}_{t}- \hat{f}^{\alpha}(t)\phi_{\alpha}(t)\,\hat{\Xi}_{t}\right]=\sum_{n=1}^{\infty}(-i)^{n}\hat{d}_{n+1}
\end{align}
where
\begin{align}\label{dLambda}
\hat{d}_{n+1}&= \sum_{\alpha_i}\hat{f}_{\alpha}(t) \int_0^t d\boldsymbol{\tau}_{n} \hat{f}_{\alpha_{1}}(\tau_1)\mcdots \hat{f}_{\alpha_{n}}(\tau_n)\nonumber\\
&\times\sum_{k=1}^{\lfloor n/2\rfloor}\sum_{\mathcal{P}(a)} \underbrace{
K_{\alpha \alpha_{1}}\!(t,\tau_{a_1})\mcdots K_{\alpha_{2k-2} \alpha_{2k-1}}\!(\tau_{a_{2k-2}},\tau_{a_{2k-1}})}_{k\, \text{elements}}\nonumber\\
&\times\underbrace{\phi_{\alpha_{2k}}(\tau_{a_{2k}})\mcdots\phi_{\alpha_{n}}(\tau_{a_{n}})}_{(n-k)\,\text{elements}}\theta_{\tau_{a_{2k}}\cdots \tau_{a_{n}}}\,.
\end{align}
Exploiting the result in Appendix B of~\cite{GasFer17} we can now rewrite the inverse map as:
\begin{align}
\hat{\Xi}^{-1}_{t}=\sum_{n=0}^{\infty}(-i)^{n}\hat{M}_{n}\,,
\end{align}
with $M_{0}=1$ and 
\begin{align}
\hat{M}_{n}= \sum_{k=1}^{n}\hat{M}_{k}\hat{\xi}_{n-k}.
\end{align}
Recollecting these results, replacing them in Eq.~\eqref{Lform},  and following the strategy in Appendix C of \cite{GasFer17}, we eventually obtain the following perturbative series for the generator $\hat{\Lambda}_{t}$:
\begin{align}
\hat{\Lambda}_{t}=\sum_{n=1}^{\infty}\hat{L}_{n}
\end{align}
whose elements $\hat{L}_{n}$ are defined by the following recursive formula 
\begin{align}
\hat{L}_{n} =\hat{d}_{n} - \sum_{k=1}^{n}\hat{L}_{k} \hat{\xi}_{n-k}\,.
\end{align}
It is important to stress that the order $n$ in the expansion terms $\hat{\xi}_{n}$, $\hat{d}_{n}$ and $\hat{L}_{n}$ denotes the power of operators $\hat{f}_{\alpha}$ displayed by them. This allows to understand the series as an expansion in the coupling strength of the interaction, which provides a useful tool for the practical analysis of non-Markovian unravellings.
Equation~\eqref{fn} instead, thought being much more elegant cannot be directly used --unless the functional derivative is known-- for explicit calculations.  
The explicit form of the $\hat{L}_n$ is still rather involved, making the evaluation of higher orders demanding. However, for a specific class of physical systems this problem can be solved and an explicit unravelling obtained.

\subsection{Bosonic quadratic Hamiltonian}
We assume that the system of interest is bosonic and its free dynamics is described by a quadratic Hamiltonian. The advantage provided by this family of systems is that they allow to apply Wick's theorem to disclose the time ordering of Eq.~\eqref{eq:lsd}. Indeed, these systems display linear Heisenberg equations of motions, and accordingly the commutators $[\hat{f}^{\alpha}(\tau),\hat{f}^{\beta}(s)]$ are c-functions. An example of system falling in this category is a damped harmonic oscillator: in this case the operators $\hat{f}$ are simply position and momentum of the oscillators, and contractions are combination of sine and cosine functions~\cite{Fer1617}. The strategy we adopt is the following: we start from Eq.~\eqref{eq:xitayl} and, instead of solving explicitly the time ordering, we keep its formal expression. Differentiating Eq.~\eqref{eq:xitayl} and exploiting the properties of the Cauchy product of two series and  we obtain:
\begin{align}\label{dotxin}
i\partial_t\hat{\Xi}_n&= \fo^j(t)\phi_j(t) \,\hat{\Xi}_{t}\\
&-i\,\hat{f}^{\alpha}(t)\mathcal{T}\left[\int_{0}^{t}\!d\tau K_{\alpha\beta}(t,\tau)\hat{f}^{\beta}(\tau)\sum_{n=0}^{\infty}\sum_{k=0}^{\infty}\cdots\right]\,\,\nonumber
\end{align}
where the dots denote the same argument of the series of Eq.~\eqref{eq:xitayl}. Note that this is just a convenient rephrasing of Eq.~\eqref{eq:lsd}. Our aim is to achieve an equation of the type~\eqref{eq22}, i.e. to  express $\partial_t\hat{\Xi}_t$ in terms of $\hat{\Xi}_{t}$. By applying Wick's theorem we find that the time ordering in the second line of Eq.~\eqref{dotxin} can be rewritten as follows:
\begin{align} \label{dotGn1}
&\mathcal{T}\left\{\int_0^t d\tau K_{\alpha\beta}(t,\tau)\hat{f}^{\beta}(\tau)\sum_{n=0}^{\infty}\sum_{k=0}^{\infty}\mcdots\right\}=\nonumber\\
&\int_0^t d\tau K_{\alpha\beta}(t,\tau)\left(\hat{f}^{\beta}(\tau)-i\int_0^t d\tau_1\,\mathfrak{C}^{\beta\alpha_{1}}(\tau,\tau_{1})\phi_{\alpha_{1}}(\tau_{1})\right)\Xi_{t}\nonumber\\
&-\mathcal{T}\left\{\int_0^t d\tau \hat{f}^{\beta}(\tau)K_{\alpha\beta}^{(2)}(t,\tau)\sum_{n=0}^{\infty}\sum_{k=0}^{\infty}\cdots\right\}
\end{align}
where $K^{(2)}$ is a suitably defined kernel, and $\mathfrak{C}^{\beta\alpha}(\tau_1,\tau)=[\hat{f}^{\beta}(\tau_1),\hat{f}^{\alpha}(\tau)]\theta_{\tau_1\tau}  $ is a Wick contraction (see Appendix \ref{app:luca}). We have been able to decompose the initial time ordering in two terms: one proportional to $\hat{\Xi}_{t}$, and one displaying a  time ordering that has the same structure as the initial one, but with a different kernel. We can then apply Eq.~\eqref{dotGn1} to this new time ordering, and iterating this procedure we obtain (see Appendix \ref{app:luca})
\begin{eqnarray}\label{Gn3}
\partial_t\hat{\Xi}_t&=&\left[-i \fo^{\alpha}(t)\phi_{\alpha}(t)-i \fo^{\alpha}(t)\int_0^td\tau \,\mathbb{K}_{\alpha\beta}(t,\tau)\right.\\
&&\hspace{0.2cm}\left.\times\left(-i\fo^{\beta}(\tau)+\!\int_0^t\!d\tau_1\, \mathfrak{C}^{\beta\alpha_{1}}(\tau,\tau_1)\phi_{\alpha_{1}}(\tau_1) \right)\!\right]\!\hat{\Xi}_{t}\,,\nonumber
\end{eqnarray}
where $\mathbb{K}=\sum_{n=1}^\infty K^{(n)}$, with $K^{(1)}=K$, and $K^{(n)}$ defined recursively through the following formula
\begin{align}\label{eq:knapp}
K_{\alpha\beta}^{(n)}(t,\tau)\!=\!\!\int_0^t\!\!\!\! d\boldsymbol{\tau}_2 K^{(n-1)}_{\alpha \alpha_{2}}(t,\tau_2)\mathfrak{C}^{\alpha_{2}\alpha_{1}}(\tau_2,\tau_1)\overline{K}_{\alpha_{1}\beta}(\tau_1,\tau)\,,
\end{align}
with $\overline{K}_{\alpha_{1}\beta}(\tau_{1},\tau)={K}_{\alpha_1\beta}(\tau_{1},\tau)+{K}_{\beta\alpha_{1}}(\tau,\tau_{1})$.
The condition for convergence of the series is:
\begin{align}\label{eq:con}
\lim_{n\to \infty}K_{\alpha\beta}^{(n)}(t,\tau)\to 0\,.
\end{align}
This equation provides the explicit non-Markovian unravelling for the family of bosonic quadratic systems, generalizing the result obtained in~\cite{BasFer09} for a specific model.
Eventually, the average dynamics associated to this equation is described by the master equation obtained in~\cite{Fer1617}.

\section{Conclusions}
We have first investigated open quantum systems with factorized initial states, and we derived the structure of the reduced CPTP map, making explicit its dependence on bath cumulants.
We then considered the class of linear SSEs with continuous stochastic processes, exploring the possibility of using them to construct CPTP dynamics. 
We found that when real stochastic process are considered the structure of the unravelling is trivial.
 On the other hand, we showed that when SSEs with complex noises are considered, the conditions such that these provide average CPTP maps are rather involved. Furthermore, we provided the condition that SSEs have to fulfil in order to unravel open quantum systems dynamics.


We eventually focused on Gaussian unravellings,  providing a new perturbative expansion that relies on a recursive equation. This expansion, though being less elegant than the known one displaying a functional derivative, allows to compute recursively the non-local term of the unravelling.
Furthermore, for the family of quadratic bosonic systems, we have been able to provide an explicit operatorial expression for the unravelling.

\section*{Acknowledgments}
 The authors are grateful to L. Diosi for spotting a flaw on a first derivation of the conditions for the non-Gaussian stochastic unraveling. The authors further thank A. Bassi for useful discussions.
GG acknowledges funding from the European Union's Horizon 2020 research and innovation programme under grant agreement No 766900. LF acknowledges funding from the Royal Society under the Newton International Fellowship No NF170345. During his stay at the University of Ljubljana, LF has been supported by the ERC Advanced Grant 694544 - OMNES.


\onecolumngrid
\appendix

\section{}
\label{app:cum}
In this Appendix we provide explicit calculation leading to Eqs.~\eqref{cuMt}-\eqref{kn0}.
We start from Eq.~\eqref{eq50}:
\begin{align}\label{eq:mapmm}
\mathcal{M}_{t}[\hat{\rho}_{\s}]\!=\!\braket{\mathcal{T}e^{-\frac{i}{2}\int_{0}^{t}\!\!d\tau \sum_\alpha [{f}_{\alpha}^{+}(\tau)\phi^{-}_{\alpha}(\tau)+{f}_{\alpha}^{-}(\tau){\phi}^{+}_{\alpha}(\tau)]}\hat{\rho}_{\s}}
\end{align}
where $f^{\pm}$  ($\phi^{\pm}$) are super-operators acting respectively on the system (environment) Hilbert space (definition in the main text), and $\braket{\cdots}$ denotes the average over a certain measurable space (in this case the trace over the environment degrees of freedom).
Exploiting Eq.~\eqref{eq:taylort} we can define the time ordered logarithm as follows:
\begin{align}
\mathcal{T}\log\left[1+f\left(\int_{0}^{t}d\tau\hat{V}_{\tau}\right)\right]= \sum_{n=1}^{\infty}\frac{(-)^{n+1}}{n}\mathcal{T}\left[f\left(\int_{0}^{t}d\tau\hat{V}_{\tau}\right)^{n}\right]
\end{align}
provided that $|f(\int_{0}^{t}\hat{V}_{\tau})|< 1$.
One can check that such time ordered logarithm satisfies the identity $\mathcal{T}\log(\mathcal{T}e^{\int_{0}^{t}d\tau \hat{V}_{\tau}})=\int_{0}^{t}d\tau \hat{V}_{\tau}$. Exploiting this in Eq.~\eqref{eq:mapmm} we find:
\begin{align}\label{logM}
\mathcal{M}_{t}\!=\!\mathcal{T}\!\exp\bigg\{\!\mathcal{T}\!\log\! \braket{\!\mathcal{T}e^{\!-\frac{i}{2}\!\int_{0}^{t}\!d\tau \!\sum_{\alpha}[f^{+}_{\alpha}(\tau)\phi^{-}_{\alpha}(\tau)+f^{-}_{\alpha}(\tau)\phi^{+}_{\alpha}(\tau)]}\!}\!\!\!\bigg\}\,.
\end{align}
The exponent of this expression can be rewritten as the following limit
\begin{align}
\hspace{-6mm}\lim_{\hspace{6mm}{\varepsilon}_{\alpha}^{\pm}(\tau)\to 0}\hspace{-0.6cm}\mathcal{T}\log\!\braket{\!\mathcal{T}\!e^{-\frac{i}{2}\!\int_{0}^{t}\! d\tau [(f^{+}_{\alpha}\!(\tau)+i\varepsilon_{\alpha}^{+}\!(\tau))\phi^{-}_{\alpha}\!(\tau)+(f^{-}_{\alpha}\!(\tau)+i{\varepsilon}_{\alpha}^{-}\!(\tau))\phi^{+}_{\alpha}(\tau)]}\!},
\end{align}
and exploiting functional calculus~\cite{functional}, one can conveniently rewrite the logarithm in Eq.~\eqref{logM} as follows
\begin{align}\label{eq:logeps}
&\mathcal{T}\log \braket{\mathcal{T}e^{-\frac{i}{2}\int_{0}^{t}d\tau \sum_{\alpha}[f^{+}_{\alpha}(\tau)\phi^{-}_{\alpha}(\tau)+f^{-}_{\alpha}(\tau)\phi^{+}_{\alpha}(\tau)]}}\nonumber\\
&\hspace{3mm}=\hspace{-3mm}\lim_{\hspace{3mm}{\varepsilon}_{\alpha}^{\pm}(\tau)\to 0}\hspace{-0.2cm}\mathcal{T}e^{ -i \int d\tau\, \sum_{\alpha}[f^{+}_{\alpha}(\tau)\delta/\delta{\varepsilon^{-}_{\alpha}(\tau)}+f^{-}_{\alpha}(\tau)\delta/\delta{\varepsilon^{+}_{\alpha}(\tau)}]} \log\braket{\mathcal{T}e^{\frac{1}{2}\int_{0}^{t} d\tau\,\sum_{\beta}[\varepsilon_{\beta}^{-}(\tau)\phi_{\beta}^{-}(\tau)+\varepsilon_{\beta}^{-}(\tau)\phi_{\beta}^{+}(\tau)]}}
\end{align}
where  $\delta / \delta \varepsilon^{\pm}_{\alpha}(\tau)$ denotes a functional derivative. This expression is convenient because it allows to evaluate independently the system and the environment contributions to the system dynamics.
One may notice indeed that, in the r.h.s. of Eq.~\eqref{eq:logeps},  the super-operators $f^{\pm}$ acting on the system are all collected in the first exponential, while the remaining degrees of freedom are displayed only by the logarithm.
Expanding the exponential function of the system super-operators $f^{\pm}_{\alpha}(\tau)$ and resolving its time ordering one obtains
\begin{align}
&\mathcal{T}\log\braket{\mathcal{T}e^{-\frac{i}{2}\int_{0}^{t} d\tau \sum_{\alpha}[(f^{+}_{\alpha}(\tau)+i\varepsilon_{\alpha}^{-}(\tau))\phi^{-}_{\alpha}(\tau)+(f^{-}_{\alpha}(\tau)+i\varepsilon_{\alpha}^{-}(\tau))\phi^{+}_{\alpha}(\tau)]} }\nonumber\\
&= \sum_{i=0}^{\infty} \frac{(-i)^{n}}{n!}\int_0^t\! d\boldsymbol{\tau}_{n}\!\mathcal{T}\bigg\{\sum_{\alpha}
\bigg[f^{+}_{\alpha}(\tau)\frac{\delta}{\delta \varepsilon^{-}_{\alpha}(\tau)}+f^{-}_{\alpha}(\tau)\frac{\delta}{\delta \varepsilon^{+}_{\alpha}(\tau)}\bigg]^{n}\bigg\}
\log\braket{\mathcal{T}e^{\frac{1}{2}\int_{0}^{t} d\tau\, \sum_{\beta}[\varepsilon_{\beta}^{-}(\tau)\phi_{\beta}^{-}(\tau)+\varepsilon_{\beta}^{-}(\tau)\phi_{\beta}^{+}(\tau)]}}\nonumber\\
&=\sum_{i=0}^{\infty} (-i)^{n}\int_0^t\! d\boldsymbol{\tau}_{n}\! \sum_{q=0}^{n}\sum_{\mathcal{P}_{q} }\!\!\sum_{\alpha_{1}\cdots \alpha_{n}}\hspace{-0.2cm}{f}_{\alpha_{1}}^{l_{1}}(\tau_{1})\mcdots{f}_{\alpha_{n}}^{l_{n}}(\tau_{n})\theta_{\tau_{1}\cdots\tau_{n}}\frac{\delta}{\delta \varepsilon_{\alpha_{1}}^{\bar{l}_{1}}(\tau_{1})}\mcdots\frac{\delta}{\delta \varepsilon_{\alpha_{n}}^{\bar{l}_{n}}(\tau_{n})}
\log \braket{\mathcal{T}e^{\frac{1}{2}\int_{0}^{t} d\tau\, \sum_{\beta}[\varepsilon_{\beta}^{+}(\tau)\phi_{\beta}^{-}(\tau)+\varepsilon_{\beta}^{-}(\tau)\phi_{\beta}^{+}(\tau)]}}\nonumber\\
\end{align}

where $\mathcal{P}_{q}$ denotes all the permutation of the indexes $k_{q} \in \{+, - \}$, such that there is a q number of minus super-operators.
Performing the limit of $\varepsilon_{\alpha}^{\pm}(\tau)\to 0$ one eventually obtains
\begin{align}\label{eq:logeexp}
&\mathcal{T}\log\braket{\mathcal{T}e^{-\frac{i}{2}\int_{0}^{t} d\tau \sum_{\alpha}(f^{+}_{\alpha}(\tau)\phi^{-}_{\alpha}(\tau)+f^{-}_{\alpha}(\tau)\phi^{+}_{\alpha}(\tau))} }\nonumber\\
&= \sum_{i=0}^{\infty}\!{(-i)^{n}}\!\int_0^t\!\! d\boldsymbol{\tau}_{n}\sum_{j=0}^{n} \sum_{\alpha_{i},\mathcal{P}_{q}}{f}_{\alpha_{1}}^{l_{1}}(\tau_{1})\!\mcdots\! {f}_{\alpha_{n}}^{l_{n}}(\tau_{n}) \widetilde{C}^{\bar{l}_{1}\cdots \bar{l}_{n}}_{\alpha_{1}\cdots\alpha_{n}}(\boldsymbol{\tau}_{n})\,,
\end{align}
where the contribution given by the auxiliary degrees of freedom is described by the ordered cumulants: 
\begin{align}\label{eq:cdc}
&\widetilde{C}^{\bar{l}_{1}\cdots \bar{l}_{n}}_{\alpha_{1}\cdots\alpha_{n}}(\boldsymbol{\tau}_{n})=\theta_{\tau_{1}\cdots\tau_{n}} \delta_{\varepsilon_{\alpha_{1},\tau_{1}}}\!\mcdots\delta_{\varepsilon_{\alpha_{n},\tau_{n}}}\log\braket{\mathcal{T} e^{\frac{1}{2}\int d\tau  \varepsilon_{\alpha,\tau}\phi^{\alpha,\tau}}}\Big|_{\varepsilon_{\alpha_{i},\tau_{i}}=0}.
\end{align}
A more explicit expression for the cumulants can be obtained by exploiting Ursell formula~\cite{ursell}, that allows to write
\begin{align}\label{cumC}
\widetilde{C}_{\alpha_{1},\cdots\alpha_{n}}^{+\cdots \bar{l}_{n}}(\boldsymbol{\tau}_{n})\!=\!\sum_{\mathbb{P}}\!\frac{(\abs{\!\mathbb{P}}\hspace{-0.1cm}-\hspace{-0.1cm} 1 \hspace{-0.05cm})!}{(\hspace{-0.4mm}-1\hspace{-0.4mm})^{\abs{\!1-\mathbb{P}}}} \hspace{-0.07cm}\prod_{P\in \mathbb{P}}\hspace{-0.15cm}\braket{\hspace{-0.07cm}\mathcal{T}\bigg(\prod_{p\in P}\frac{{\phi}_{\alpha_{p}}^{\bar{l}_{p}}(\tau_{p})}{2}\bigg)\!\!}\!\theta_{\tau_{1}\dots\tau_{n}}
\end{align}
where  $\mathbb{P}$ is a partition of $\{1,\dots,n\}$ with cardinality $\abs{\mathbb{P}}$, $P\in \mathbb{P}$ is a set of the partition $\mathbb{P}$, and $p$ one element of the set $P$. 
Exploiting Eq.~\eqref{eq:logeexp} in Eq.~\eqref{eq:mapmm} and introducing the new symbol 
\begin{align}\label{kapp}
\tilde{k}_{n}(\boldsymbol{\tau}_{n})=\!\! \sum_{\alpha_{i},\mathcal{P}_{q}}{f}_{\alpha_{1}}^{l_{1}}(\tau_{1})\!\mcdots\! {f}_{\alpha_{n}}^{l_{n}}(\tau_{n}) \widetilde{C}^{\bar{l}_{1}\cdots \bar{l}_{n}}_{\alpha_{1}\cdots\alpha_{n}}(\boldsymbol{\tau}_{n}),
\end{align}
 we eventually obtain Eq.~\eqref{kn0} of the main text.
 
 We stress that the derivation provided in this Appendix holds also if $\phi^\pm$ in Eq.~\eqref{eq:mapmm} are stochastic processes and  $\braket{\cdots}$ denotes the stochastic average. The final result is Eq.~\eqref{kapp} where $\widetilde{k}\rightarrow k$ and $\widetilde{C}\rightarrow C$, i.e. Eq.~\eqref{kn}. It is important to remark that in general $\widetilde{C}$ and $C$ differ because quantum trace and stochastic average have different properties.

\section{}
\label{app:tr}
In this Appendix we provide calculations  leading to Eq.~\eqref{eq:Aop}.
The map $\mathcal{N}_{t}$ in Eq.~\eqref{mtilde} is TP if $\Tr[\mathcal{N}_{t}\hat{\rho}_{\s}]=\Tr[\hat{\rho}_{\s}]$ or equivalently $\partial_{t}\Tr[\mathcal{N}_{t}\hat{\rho}_{\s}]=0$.
Performing the trace and taking the time derivative we get
\begin{align}
i\partial_{t}\Tr\left\{\mathcal{N}_{t}[\hat{\rho}_{\s}]\right\}=&\Tr\Bigg\{\,\mathcal{T}\bigg[\int_0^t d\boldsymbol{\tau}_n\Bigg(\sum_{n=1}^{\infty}\!\sum_{q=1}^{n}\!\sum_{\alpha_i,\mathcal{P}_{q}}
(-i)^{n}f_{\alpha_{0}}^{+}(t)f^{l_{1}}_{\alpha_1}(\tau_{1})\mcdots f^{l_{n}}_{\alpha_n}(\tau_{n})\,C^{- \bar{l}_{1}\cdots \bar{l}_{n}}_{\alpha_0\cdots\alpha_n}(\boldsymbol{\tau}_n)-iA^{+}(\boldsymbol{\tau}_{n})\Bigg)\nonumber\\
&\times\exp\left[\sum_{n=1}^{\infty} \int_0^t d\boldsymbol{\tau}_{n}\Big((-i)^{n}k_{n}(\boldsymbol{\tau}_{n})-\mathcal{O}_{n}(\boldsymbol{\tau}_{n})\Big)\right]\hat{\rho}_{\s}\Bigg\}\,.
\end{align}
Exploiting the cyclicity of the trace we can rewrite the equation as
\begin{align}
i\partial_{t}\Tr\left\{\mathcal{N}_{t}[\hat{\rho}_{\s}]\right\}=&\Tr\Bigg\{\,\mathcal{T}\bigg[\int_0^t d\boldsymbol{\tau}_n\Bigg(\sum_{n=1}^{\infty}\!\sum_{q=1}^{n}\!\sum_{\alpha_i,\mathcal{P}_{q}}(-i)^{n}\hat{f}_{\alpha_{0}}(t)\harpl{f}^{l_{1}}_{\alpha_1}(\tau_{1})\mcdots \harpl{f}^{l_{n}}_{\alpha_n}(\tau_{n})\,C^{- \bar{l}_{1}\cdots \bar{l}_{n}}_{\alpha_0\cdots\alpha_n}(\boldsymbol{\tau}_n)-i\hat{A}(\boldsymbol{\tau}_{n})\Bigg)\nonumber\\
&\times\exp\left[\sum_{n=1}^{\infty} \int_0^t d\boldsymbol{\tau}_{n}\Big((-i)^{n}\harpl{k}_{n}(\boldsymbol{\tau}_{n})-\harpl{\mathcal{O}}_{n}(\boldsymbol{\tau}_{n})\Big)\right]\hat{\rho}_{\s}\Bigg\}=0;
\end{align}
where the super-operator $\harpl{k}(\boldsymbol{\tau_{n}})$ is ${k}(\boldsymbol{\tau_{n}})$ where all the ${f}^{l_{i}}$ have been replaced by $\harpl{f}^{l_{i}}$. From the above equation we immediately obtain Eq.~\eqref{eq:Aop} for the operator $\hat{A}(\boldsymbol{\tau}_{n})$.
This condition is rather cumbersome, and one might argue that a simpler one can be obtained by expanding the operators $\hat{O}_n(\boldsymbol{\tau}_{n})$ over the set $\{\hat{f}_{\alpha}\}$ as follows:
\begin{align}\label{opn}
\hat{O}_n(\boldsymbol{\tau}_{n})=\sum_{\alpha_i} \hat{f}_{\alpha_1}(\tau_1)\mcdots\hat{f}_{\alpha_n}(\tau_n)\,K_{\alpha_1\cdots\alpha_n}(\boldsymbol{\tau}_{n})\,,
\end{align}
where the $K_{\alpha_1\cdots\alpha_n}$ are complex kernels. 
However, we now show that a decomposition of this kind works only in the Gaussian case, while it is not instructive for noises with cumulants higher than the second.
With the choice~\eqref{opn}, the stochastic map~\eqref{xi} generates the average dynamics~\eqref{mtilde}
with (see Appendix~\ref{app:C})
\begin{align}\label{eq:on}
\mathcal{O}_{n}(\boldsymbol{\tau}_n)&=\sum_{q=0}^{n}\sum_{\alpha_i,\mathcal{P}_{q}}f^{l_{1}}_{\alpha_{1}}(\tau_1)\mcdots f^{l_{n}}_{\alpha_{n}}(\tau_n)\,K^{\diamond}_{\alpha_{1}\cdots \alpha_{n}}\!(\boldsymbol{\tau}_n)\,,
\end{align}
where $\diamond=+$ when $q$ is even, $\diamond=-$ when $q$ is odd, and we have introduced $K^\pm=(K\pm K^*)/2$.
One notices that the structure of the super-operators $\mathcal{O}_n(\boldsymbol{\tau}_{n})$ is close to that of $k_n(\boldsymbol{\tau}_{n})$. The relevant difference is that, unlike $C$, $K$ depends only on the number $q$ of plus signs of the array $({l}_{1},\cdots {l}_n)$, not on the array itself (note different superscripts). Substituting Eqs.~\eqref{kn} and \eqref{eq:on} in Eq.~\eqref{mtilde} one finds
\begin{align}\label{eq:efmmap}
\mathcal{N}_{t}[\hat{\rho}_{\s}] &= \mathcal{T}\exp\Bigg[\int_0^td\boldsymbol{\tau}_n\sum_{n=1}^{\infty}\sum_{q=1}^{n}\sum_{\alpha_i,\mathcal{P}_{q}}(-i)^nf^{l_{1}}_{\alpha_{1}}(\tau_1)\mcdots f^{l_{n}}_{\alpha_{n}}(\tau_n)\left(C^{\bar{l}_{1}\cdots \bar{l}_{n}}_{\alpha_{1}\cdots \alpha_{n}}\!(\boldsymbol{\tau}_n)-K^{\diamond}_{\alpha_{1}\cdots \alpha_{n}}\!(\boldsymbol{\tau}_n)\right)\!\!\Bigg]\hat{\rho}_{\s}\,.
\end{align}
By performing the trace of this equation, one finds that the averaged map $\mathcal{N}_t$ is TP only if the following conditions are satisfied: $\forall n \in \mathbb{N},\, q\leq n$
\begin{align}\label{eq:sist}
C^{-\bar{l}_{2}\cdots \bar{l}_{n}}_{\alpha_{1}\!\cdots \alpha_{n}}\!(\boldsymbol{\tau}_n)\!-\!K^{+}_{\alpha_{1}\!\cdots \alpha_{n}}\!(\boldsymbol{\tau}_n)&\!=\!0\, ,\quad   q\,\, \mathrm{even}\nonumber\\
C^{-\bar{l}_{2}\cdots \bar{l}_{n}}_{\alpha_{1}\!\cdots \alpha_{n}}\!(\boldsymbol{\tau}_n)\!-\!K^{-}_{\alpha_{1}\!\cdots \alpha_{n}}\!(\boldsymbol{\tau}_n)&\!=\!0\, ,\quad   q\,\, \mathrm{odd}.
\end{align}
We recall that $\bar{l}_{1}= -$ because of the trace operation (see Eq.~\eqref{eq:tr}). 
Since the identities of Eq.~\eqref{eq:sist} must hold for any array $(l_{2}, \cdots l_{n})$, they represent a system of $2^{n-1}$ independent equations. However, at any order $n$ there are only two free variables: $K^{+}_{\alpha_{1}\cdots \alpha_{n}}\!(\boldsymbol{\tau}_n)$ and $K^{-}_{\alpha_{1}\cdots \alpha_{n}}\!(\boldsymbol{\tau}_n)$. This is a consequence of the fact that $C$ depends on the array $(-,\bar{l}_{2},\cdots \bar{l}_{n})$, while $K$ depends only on the number of minus signs in such an array.
Accordingly, we can conclude that the system~\eqref{eq:sist} admits solution only for $n\le 2$, and that the decomposition~\eqref{opn} allows to obtain the condition for a TP map only in the Gaussian case.

\section{}
\label{app:C}

In this Appendix we derive Eqs.~\eqref{eq:on} for the modified map $\mathcal{N}_t$.
It is convenient to introduce the left-right formalism denoting by a superscript L (R) the operators acting on $\hat{\rho}$ from the left (right)~\cite{LR}.
The map $\mathcal{N}_{t}=\langle\Xi_{t}\rho\Xi_{t}^{\dagger}\rangle$, defined in the main text, can be then written as:
\begin{align}\label{NappB}
\mathcal{N}_{t}&=\Bigg\langle\!\mathcal{T}\!\exp\bigg\{\sum_{n=1}^{\infty}(-i)^{n}\int_{0}^{t}d\boldsymbol{\tau}_{n} O^{L}_{n}(\boldsymbol{\tau}_{n})+(-)^{n}O^{R}_{n}(\boldsymbol{\tau}_{n})^{\dagger}{-i\int_{0}^{t}d\tau\sum_{\alpha}[f_{\alpha}^{L}(\tau)\phi^{L}_{\alpha}(\tau)+f^{R}_{\alpha}(\tau)\phi^{R}_{\alpha}(\tau)]}\bigg\}\Bigg\rangle\,.
\end{align}
We introduce the super-operator $\mathcal{O}_{n}(\boldsymbol{\tau}_{n})$:
 \begin{align}
\mathcal{O}_{n}(\boldsymbol{\tau}_{n})= O^{L}_{n}(\boldsymbol{\tau}_{n})+(-)^{n}O^{R}_{n}(\boldsymbol{\tau}_{n})^{\dagger}\,,
\end{align}
which, making use of Eq.~\eqref{opn} and of the identity $K^{\pm}=(K\pm K^{*})/2$, can be rewritten as follows:
\begin{align}\label{eq:Otnapp}
\mathcal{O}_{n}(\boldsymbol{\tau}_{n})&=\nonumber\\ 
&\hspace{-0.5cm}[{f}_{\alpha_{1}}^{L}\!(\tau_{1})\!\mcdots\!{f}^{L}_{\alpha_{n}}\!(\tau_{n})
 +(-)^{n} {f}_{\alpha_{1}}^{R}\!(\tau_{1})\!\mcdots\!{f}^{R}_{\alpha_{n}}\!(\tau_{n})]{K}^{+}_{\alpha_{1}\!,\cdots,\alpha_{n}}\!(\boldsymbol{\tau}_{n})\!+\![{f}_{\alpha_{1}}^{L}\!(\tau_{1})\!\mcdots\!{f}^{L}_{\alpha_{n}}\!(\tau_{n})
 -(-)^{n} {f}_{\alpha_{1}}^{R}\!(\tau_{1})\!\mcdots\!{f}^{R}_{\alpha_{n}}\!(\tau_{n})]{K}^{-}_{\alpha_{1}\!,\cdots,\alpha_{n}}\!(\boldsymbol{\tau}_{n})\,.
\end{align}
Further rewriting $f^{R/L}$ in terms of $f^{\pm}= (f^{L}\pm f^{R})/2$, we obtain
\begin{align}\label{eq:Otn}
&\mathcal{O}_{n}(\boldsymbol{\tau}_{n})=\big\{[{f}_{\alpha_{1}}^{+}\!(\tau_{1})+{f}_{\alpha_{1}}^{-}\!(\tau_{1})]\mcdots[{f}_{\alpha_{n}}^{+}\!(\tau_{n})+{f}_{\alpha_{n}}^{-}\!(\tau_{n})]+
(-)^{n}[{f}_{\alpha_{1}}^{+}\!(\tau_{1})-{f}_{\alpha_{1}}^{-}\!(\tau_{1})]\mcdots[{f}_{\alpha_{n}}^{+}\!(\tau_{n})-{f}_{\alpha_{n}}^{-}\!(\tau_{n}]\big\}{K}^{+}_{\alpha_{1}\!,\cdots\!,\alpha_{n}}\!(\boldsymbol{\tau}_{n})\nonumber\\
&\hspace{1.21cm}+\big\{[{f}_{\alpha_{1}}^{+}\!(\tau_{1})+{f}_{\alpha_{1}}^{-}\!(\tau_{1})]\mcdots[{f}_{\alpha_{n}}^{+}\!(\tau_{n})+{f}_{\alpha_{n}}^{-}\!(\tau_{n})]-
(-)^{n}[{f}_{\alpha_{1}}^{+}\!(\tau_{1})-{f}_{\alpha_{1}}^{-}\!(\tau_{1})]\mcdots[{f}_{\alpha_{n}}^{+}\!(\tau_{n})-{f}_{\alpha_{n}}^{-}\!(\tau_{n}]\big\}{K}^{-}_{\alpha_{1}\!,\cdots\!,\alpha_{n}}\!(\boldsymbol{\tau}_{n}) \nonumber\\
&\!\!\!=\!\sum_{j=1}^{n}\!\sum_{\alpha_{i},\mathcal{P}_{q}}\!\!\bigg\{\!\!\big[f^{l_{1}}_{\alpha_{1}}\!(\tau_{1})\!\mcdots\!\! f^{l_{n}}_{\alpha_{n}}\!(\tau_{n})\!+\!(-)^{q}\!f^{l_{1}}_{\alpha_{1}}\!(\tau_{1})\!\mcdots\! f^{l_{n}}_{\alpha_{n}}\!(\tau_{n})\big]\!{K}^{+}_{\alpha_{1}\!,\cdots\!,\alpha_{n}}\!\!(\boldsymbol{\tau}_{n})
\!+\!\big[f^{l_{1}}_{\alpha_{1}}\!(\tau_{1})\!\mcdots \!\!f^{l_{n}}_{\alpha_{n}}\!(\tau_{n})\!-\!(-)^{q}\!f^{l_{1}}_{\alpha_{1}}\!(\tau_{1})\!\mcdots\! f^{l_{n}}_{\alpha_{n}}\!(\tau_{n})\big]\!{K}^{-}_{\alpha_{1}\!,\cdots\!,\alpha_{n}}\!\!(\boldsymbol{\tau}_{n})\!\!\bigg\}\,,\nonumber\\
\end{align}

where $\mathcal{P}_{q}$ is the permutation of the indexes $l_{i}$ such that $q$ is the number of plus signs in the array $({l}_1,\cdots, {l}_n)$, and $\alpha_{i}$ denotes the sum over all $\{\alpha_{1},\cdots,\alpha_{n}\}$.
The last line of this equation shows that the terms proportional to $K^{+}$ ($K^{-}$) with odd (even) $q$  are zero, allowing to rewrite the $\mathcal{O}_{n}(\boldsymbol{\tau}_{n})$ as follows
\begin{align}\label{eq:onao}
\mathcal{O}_{n}(\boldsymbol{\tau}_n)&=\sum_{q=0}^{n}\sum_{\alpha_i,\mathcal{P}_{q}}f^{l_{1}}_{\alpha_{1}}(\tau_1)\cdots f^{l_{n}}_{\alpha_{n}}(\tau_n)\,K^{\diamond}_{\alpha_{1}\cdots \alpha_{n}}\!(\boldsymbol{\tau}_n)\,,
\end{align}
where $\diamond=+$ when $q$ is even, $\diamond=-$ when $q$ is odd, providing us with Eq.~\eqref{eq:on}.

\section{}
\label{app:B}
In this Appendix we provide explicit calculation for the series expansion of Eqs.\eqref{qn}-\eqref{dLambda}.
Since our aim is to provide a series in powers of interaction operators $\hat{f}^\alpha$, we need to rearrange the series in Eq.~\eqref{eq:xitayl} because this contains terms with powers of interaction operators in the range $[n, 2n]$.
The strategy is to disentangle the two series in Eq.~\eqref{eq:xitayl} by exploiting the properties of the Cauchy product of two series:
\begin{align}\label{eq:xiapp0}
\hat{\Xi}_{t}=
\mathcal{T}\sum_{n=0}^{\infty}\sum_{k=0}^{\infty}\frac{1}{n!k!}\left(-i\int_0^t d {\tau}\hat{f}^{\alpha}(\tau)\phi_{\alpha}(\tau)\right)^{n}
\left(-\int_0^t d\boldsymbol{\tau}_{2}\hat{f}^{\alpha_{1}}(\tau_{1})\hat{f}^{\alpha_{2}}(\tau_{2})K_{\alpha_{1}\alpha_{2}}(\boldsymbol{\tau}_{2})\right)^{k}.
\end{align}
Notice that, at each increment of $k$ the order of the $\hat{f}^\alpha$ increases by two, while at each increment of $n$ it increases just by one. 
We replace $k \to \lfloor k/2\rfloor$ in order to have the same order of operators $\hat{f}^\alpha$ in both series and obtain:
\begin{align}
\hat{\Xi}_{t}=&
\sum_{n=0}^{\infty}\sum_{k=0}^{\infty}\frac{1}{n!\lfloor k/2\rfloor!}\mathcal{T}\bigg\{\left(-i\int_0^t d {\tau}\hat{f}^{\alpha}(\tau)\phi_{\alpha}(\tau)\right)^{n}\frac{1}{2}\left(-\int_0^t d\boldsymbol{\tau}_{2}\hat{f}^{\alpha_{1}}(\tau_{1})\hat{f}^{\alpha_{2}}(\tau_{2})K_{\alpha_{1}\alpha_{2}}(\boldsymbol{\tau}_{2})\right)^{\lfloor k/2\rfloor}\bigg\}\,.
\end{align}
The extra factor one half appearing in the second line of this equation is needed to compensate the extra contributions given by the introduction of $\lfloor k/2 \rfloor$, that counts twice the terms of the series ($\lfloor (2k+1)/2\rfloor=\lfloor (2k) / 2\rfloor $).
We can now exploit again the properties of the Cauchy product in order to entangle back the series, obtaining
\begin{align}\label{eq:xiapp}
\hat{\Xi}_t=&\sum_{n=0}^{\infty}(-i)^{n}\hat{\xi}_{t}
\end{align}
where  
\begin{align}
\hat{\xi}_{t}=\!\sum_{k=0}^{n}\frac{1}{(n-\lfloor k/2\rfloor)!\lfloor k/2\rfloor!}\mathcal{T}\bigg\{\left(\int_0^t d {\tau}\hat{f}^{\alpha}(\tau)\phi_{\alpha}(\tau)\right)^{n-k}\frac{1}{2}\left(\int_0^t d\boldsymbol{\tau}_{2}\hat{f}^{\alpha_{1}}(\tau_{1})\hat{f}^{\alpha_{2}}(\tau_{2})K_{\alpha_{1}\alpha_{2}}(\boldsymbol{\tau}_{2})\right)^{\lfloor k/2\rfloor}\!\bigg\}.
\end{align}
This series is such that the order $n$ coincides with the power of operators $\hat{f}^\alpha$.
In order to make this fact more evident we replace $k\to 2k$, to get
\begin{align}
\hat{\xi}_{t}=&
\mathcal{T}\sum_{n=0}^{\infty}\sum_{k=0}^{\lfloor n/2 \rfloor}\frac{1}{(n-2k)!k!}\mathcal{T}\bigg\{\!\!\left(\!\!-i\!\int_0^t\! \!d {\tau}\hat{f}^{\alpha}(\tau)\phi_{\alpha}(\tau)\right)^{n-2k}\hspace{-1.5mm}\frac{1}{2}\left(-\!\int_0^t\!\! d\boldsymbol{\tau}_{2}\hat{f}^{\alpha_{1}}(\tau_{1})\hat{f}^{\alpha_{2}}(\tau_{2})K_{\alpha_{1}\alpha_{2}}(\boldsymbol{\tau}_{2})\right)^{k}\bigg\}.
\end{align}
This equation shows that at each increment of $k$  one removes two terms of the type $\int_0^t\! \!d {\tau}\hat{f}^{\alpha}(\tau)\phi_{\alpha}(\tau)$ and adds one term of the type $\!\int_0^t\!\! d\boldsymbol{\tau}_{2}\hat{f}^{\alpha_{1}}(\tau_{1})\hat{f}^{\alpha_{2}}(\tau_{2})K_{\alpha_{1}\alpha_{2}}(\boldsymbol{\tau_{2}})$. In order to obtain Eq.~\eqref{qn} one only needs to make the time ordering explicit by conditioning the integrals with unit step functions. Doing so we  eventually obtain Eq.~\eqref{qn} in the main text, i.e.
\begin{align}
\hat{\xi}_{n}\!\!=\!\!\int_0^t d\boldsymbol{\tau}_{n}\!\! \sum_{\alpha_i}\hat{f}_{\alpha_{1}}(\tau_1)\mcdots \hat{f}_{\alpha_{n}}(\tau_n)\!\!\sum_{k=1}^{\lfloor n/2\rfloor}\!\!\sum_{\mathcal{P}_{a_i}}K_{\alpha_{1} \alpha_{2}}\!(\tau_{a_1},\tau_{a_2})\mcdots K_{\alpha_{2k-1} \alpha_{2k}}\!(\tau_{a_{2k-1}},\tau_{a_{2k}})\phi_{\alpha_{2k+1}}(\tau_{a_{2k+1}})\mcdots\phi_{\alpha_{n}}(\tau_{a_{n}})\theta_{\tau_{a_{2k+1}}\cdots\tau_{a_{n}}}\,.
\end{align}

In order to find Eq.~\eqref{lambdaxi}, we replace Eq.~\eqref{eq:xiapp0} in Eq.~\eqref{Lform} obtaining
\begin{align}\label{eq:xiapp1}
&\left[i\,\partial_t\hat{\Xi}_{t}- \hat{f}^{\alpha}(t)\phi_{\alpha}(t)\,\hat{\Xi}_{t}\right]=\nonumber\\
&\hspace{0.9cm}\mathcal{T}\bigg\{-i\int_0^t d\tau\hat{f}^{\alpha}(t)\hat{f}^{\beta}(\tau)K_{\alpha\beta}(t,\tau)\sum_{n=0}^{\infty}\sum_{k=0}^{\infty}\frac{1}{n!k!}\left(-i\int_0^t d {\tau}\hat{f}^{\alpha}(\tau)\phi_{\alpha}(\tau)\right)^{n}\left(-\int_0^t d\boldsymbol{\tau}_{2}\hat{f}^{\alpha_{1}}(\tau_{1})\hat{f}^{\alpha_{2}}(\tau_{2})K_{\alpha_{1}\alpha_{2}}(\boldsymbol{\tau}_{2})\right)^{k}\bigg\}.
\end{align}
Following the steps that lead to Eq.~\eqref{eq:xiapp} one obtains Eq.~\eqref{dLambda}.

\section{}
\label{app:luca}
In this Appendix we show how one can obtain Eq.~\eqref{Gn3} by applying Wick's theorem to Eq.~\eqref{dotxin}. 
As preliminary steps we separately study the following two cases:
\begin{align}\label{eq:1}
\mathcal{T} (\hat{f}(\tau_{0})e^{-i\int_0^t d {\tau}\hat{f}(\tau)\phi(\tau)})=\mathcal{T}\left\{ \hat{f}(\tau_{0})\sum_{n=0}^{\infty}\frac{1}{n!}\left(-i\int_0^t d {\tau}\hat{f}(\tau)\phi(\tau)\right)^{n}\right\}
\end{align}
and 
\begin{align}\label{eq:2}
\int_0^t \!d\tau_{0}\mathcal{T}\!\left\{K(t,\tau_{0}) \hat{f}(\tau_{0})e^{-\int_0^t d\boldsymbol{\tau}_{2}\hat{f}(\tau_{1})\hat{f}(\tau_{2})K(\boldsymbol{\tau}_{2})}\right\}=
\mathcal{T}\left\{\int_0^t d\tau_{0}K(t,\tau_{0}) \hat{f}(\tau_{0})\sum_{n=0}^{\infty}\frac{1}{n!}\left(-\int_0^t d\boldsymbol{\tau}_{2}\hat{f}(\tau_{1})\hat{f}(\tau_{2})K(\boldsymbol{\tau}_{2})\right)^{n}\right\}
\end{align}
where for the sake of simplicity we have dropped the indexes $\alpha_{i}$. 
Let us start with Eq.~\eqref{eq:1}.  Each term of the equation is composed by a string of $n$ integrated operators:
\begin{align}
\int_0^t  d\boldsymbol{\tau}_{n}\mathcal{T}[\hat{f}(\tau_{0})\hat{f}(\tau_{1})\mcdots \hat{f}(\tau_{n})].
\end{align}
Applying Wick's theorem one obtains:
\begin{align}\label{eq:st2}
&\int_0^t\!\! d\boldsymbol{\tau}_{n}\mathcal{T}[\hat{f}(\tau_{0})\hat{f}(\tau_{1})\mcdots \hat{f}(\tau_{n})]=\!\!\int_0^t \!\!d\boldsymbol{\tau}_{n} \hat{f}(\tau_{0})\mathcal{T}[\hat{f}(\tau_{1})\mcdots \hat{f}(\tau_{n})]+\sum_{i=1}^{n}\!\int_0^t\!\! d\boldsymbol{\tau}_{n}\mathfrak{C}(\tau_{0},\tau_{i})\mathcal{T}[\hat{f}(\tau_{1})\mcdots\hat{f}(\tau_{i-1}),\hat{f}(\tau_{i+1})\mcdots \hat{f}(\tau_{n})]\nonumber\\
&=\int_0^t\!\! d\boldsymbol{\tau}_{n} \hat{f}(\tau_{0})\mathcal{T}[\hat{f}(\tau_{1})\mcdots \hat{f}(\tau_{n})]+n\int_0^t d\boldsymbol{\tau}_{n}\mathfrak{C}(\tau_{0},\tau_{1})\mathcal{T}[\hat{f}(\tau_{2})\mcdots \hat{f}(\tau_{n})]\,,\end{align}
where $\mathfrak{C}(\tau_0,\tau_1)\equiv[\fo(\tau_1),\fo(\tau_0)]\theta_{\tau_1,\tau_0}$ is a Wick contraction.
Exploiting Eq.~\eqref{eq:st2} in Eq.~\eqref{eq:1} we are able to extract the operator $\hat{f}(\tau_{0})$ from the time ordering, at the cost of introducing an extra term proportional to the contraction: 
\begin{align}\label{eq:wseap}
\mathcal{T} (\hat{f}(\tau_{0})e^{-i\int_0^t d {\tau}\hat{f}^{\alpha}(\tau)\phi_{\alpha}(\tau)})
&=\left(\hat{f}(\tau_{0})-i\int_0^t d\tau_{1}\mathfrak{C}(\tau_{0},\tau_{1})\right)
\sum_{n=0}^{\infty}\frac{1}{n!}\mathcal{T}\left(-i\int_0^t d {\tau}\hat{f}^{\alpha}(\tau)\phi_{\alpha}(\tau)\right)^{n}\nonumber\\
&=\left(\hat{f}(\tau_{0})-i\int_0^t d\tau_{1}\mathfrak{C}(\tau_{0},\tau_{1})\right)\mathcal{T} e^{-i\int_0^t d {\tau}\hat{f}^{\alpha}(\tau)\phi_{\alpha}(\tau)}
\end{align}
The next step is to show how one can extract the term $\hat{f}(\tau)$ from the time ordering in Eq.~\eqref{eq:2}.
Applying Wick's theorem for the $n$-th term of the series one gets: 
\begin{align}
&\mathcal{T}\left\{\int_0^t d\tau_{0} K(t,\tau_{0})\hat{f}(\tau_{0})\frac{1}{n!}\left( -\int_0^t d\boldsymbol{\tau}_{2}\hat{f}(\tau_{1})\hat{f}(\tau_{2})K(\boldsymbol{\tau}_{2})\right)^{n}\right\}=\int_0^t d\tau_{0}K(t,\tau_{0})\hat{f}(\tau_{0})\mathcal{T}\left\{\frac{1}{n!}\left( -\int_0^t d\boldsymbol{\tau}_{2}\hat{f}(\tau_{1})\hat{f}(\tau_{2})K(\boldsymbol{\tau}_{2})\right)^{n}\right\}\nonumber\\
&-\int_0^t d\tau_{0}K(t,\tau_{0})\mathcal{T}\left\{\int_0^t d\boldsymbol{\tau}_{2}[\mathfrak{C}(\tau_{0},\tau_{1})\hat{f}(\tau_{2})+\mathfrak{C}(\tau_{0},\tau_{2})\hat{f}(\tau_{1})]K(\boldsymbol{\tau}_{2})\frac{1}{(n-1)!}\left( -\int_0^t d\boldsymbol{\tau}_{2}\hat{f}(\tau_{1})\hat{f}(\tau_{2})K(\boldsymbol{\tau}_{2})\right)^{n-1}\right\}\nonumber\\
&=\int_0^t d\tau_{0}K(t,\tau_{0})\hat{f}(\tau_{0})\mathcal{T}\left\{\frac{1}{n!}\left( -\int_0^t d\boldsymbol{\tau}_{2}\hat{f}(\tau_{1})\hat{f}(\tau_{2})K(\boldsymbol{\tau}_{2})\right)^{n}\right\}\nonumber\\
&-\int_0^t d\tau_{0}\int_0^t d\boldsymbol{\tau}_{2}K(t,\tau_{0})\mathfrak{C}(\tau_{0},\tau_{1})\mathcal{T}\left\{\overline{K}(\tau_{1},\tau_{2})\hat{f}(\tau_{2})\frac{1}{(n-1)!}\left( -\int_0^t d\boldsymbol{\tau}_{2}\hat{f}(\tau_{1})\hat{f}(\tau_{2})K(\boldsymbol{\tau}_{2})\right)^{n-1}\right\}
\end{align}
where $\overline{K}(\tau_{1},\tau_{2})=K(\tau_{1},\tau_{2})+ K(\tau_{2},\tau_{1})$. 
Applying this result to Eq.~\eqref{eq:2} one obtains:
\begin{align}\label{this}
&\mathcal{T}\left\{\hat{f}(t)\int_0^t d\tau K(t,\tau)\hat{f}(\tau)e^{-\!\int_0^t \!d\boldsymbol{\tau}_{2}\hat{f}(\tau_{1})\hat{f}(\tau_{2})K(\boldsymbol{\tau}_{2})}\right\}=-\mathcal{T}\left\{\int_0^t d\tau \hat{f}(\tau)K^{(2)}(t,\tau)e^{-\!\int_0^t \!d\boldsymbol{\tau}_{2}\hat{f}(\tau_{1})\hat{f}(\tau_{2})K(\boldsymbol{\tau}_{2})}\right\}
\nonumber\\
&\hspace{4cm}+\hat{f}(t)\int_0^t d\tau_{1} K(t,\tau_{1})\left(\hat{f}(\tau_{1})-i\int_0^t d\tau\mathfrak{C}(\tau_{1},\tau)\phi(\tau)\right)\mathcal{T}e^{-\!\int_0^t \!d\boldsymbol{\tau}_{2}\hat{f}(\tau_{1})\hat{f}(\tau_{2})K(\boldsymbol{\tau}_{2})}\,.
\end{align}
As one can see, unlike the previous case, in this specific case we are not able to close the equation at the first step because we obtain an extra term. However, such a term has the same structure as the one in the l.h.s. of Eq.~\eqref{this}, allowing to re-apply this equation to it. Iterating this procedure we obtain
\begin{align}\label{eq:knin}
&\mathcal{T}\left\{ K(t,\tau)\hat{f}(\tau)e^{-\!\int_0^t \!d\boldsymbol{\tau}_{2}\hat{f}(\tau_{1})\hat{f}(\tau_{2})K(\boldsymbol{\tau}_{2})}\right\}=\sum_{n=1}^{\infty}\int_0^t d\tau_{1}K^{(n)}(t,\tau_{1})\hat{f}(\tau_{1})\mathcal{T}e^{-\!\int_0^t \!d\boldsymbol{\tau}_{2}\hat{f}(\tau_{1})\hat{f}(\tau_{2})K(\boldsymbol{\tau}_{2})}\nonumber\\
&\hspace{6cm}-\mathcal{T}\left\{\int_0^t d\tau \hat{f}(\tau)K^{(\infty)}(t,\tau)e^{-\!\int_0^t \!d\boldsymbol{\tau}_{2}\hat{f}(\tau_{1})\hat{f}(\tau_{2})K(\boldsymbol{\tau}_{2})}\right\}.
\end{align}
with $K^{(1)}=K$ and 
\begin{align}
K^{(n)}(t,\tau)=\!\!\int_0^t\!\!\!d\boldsymbol{\tau}_2 K^{(n-1)}(t,\tau_2)\mathfrak{C}(\tau_2,\tau_1)\overline{K}(\tau_1,\tau)\,.
\end{align}
Eq.~\eqref{eq:knin} shows that one can extract the term $\hat{f}(\tau)$ from the time ordering only if 
\begin{align}
\lim_{n\to \infty}K^{(n)}=0\,
\end{align}
(otherwise an extra term would be left, not allowing to close the procedure).
This condition can be understood as a constraint on the convergence of the perturbative series.
With Eqs.~\eqref{eq:wseap} and \eqref{eq:knin} in hand it is straightforward to obtain Eq.~\eqref{Gn3} from Eq.~\eqref{dotGn1}.

\twocolumngrid
\bibliographystyle{ieeetr}

\end{document}